\documentclass[aps,prb,superscriptaddress,showpacs]{revtex4}

\usepackage{amsmath}
\usepackage{amsfonts}
\usepackage{bm}

\ifx\pdfoutput\undefined
        \usepackage{graphicx}
\else
        \usepackage[pdftex]{graphicx}
\fi

\begin{document}

\title{Pulse Propagation in Resonant  Tunneling} 

\author{Ulrich Wulf}
\affiliation{Technische Universit\"at Cottbus, 
Lehrstuhl f\"ur Theoretische Physik and IHP/BTU Joint Lab, Postfach
101344,
              03013 Cottbus, Germany }

\author{V. V. Skalozub}
\affiliation{Dniepropetrovsk National University, Dniepropetrovsk 49050,
Ukraine}

\date{\today}

\begin{abstract}
We consider the analytically solvable model of a Gaussian pulse
tunneling through a transmission resonance with a general
Fano characteristic.
It is demonstrated that the transmitted pulse contains enough
information to determine uniquely all parameters defining the
Fano resonance. This is in contrast to the measurement
of the static conductance. Our analytical model is in agreement
with numerical data published recently for the limit of a
Breit-Wigner resonance.
We identify two opposite  pulse propagation regimes:
if the resonance is broad compared to  the energetic width of
the incident Gaussian pulse
 a weakly deformed and slightly delayed transmitted Gaussian pulse is found.
In the opposite limit of a narrow resonance the dying out
of the transmitted pulse is dominated by the slow exponential
decay characteristic of a quasi-bound state with a long life time.
In this regime we find characteristic interference oscillations.
\end{abstract}


\maketitle 

\section{introduction}
In a number of semiconductor nanosystems at low enough temperatures
the ballistic transport properties 
are dominated by a single resonance.
Well known examples  are the
double barrier resonant tunneling
diode and quantum dots.  
Depending on the coupling of the dot to the contacts,
one observes in the conductance of the latter systems
Coulomb blockade oscillations, the Kondo effect,
or Fano resonances \cite{goe1,goe2,zach}. In  the Fano 
regime strongly asymmetric
conductance peaks as well as  anti-resonances are observed \cite{goe2,fuehn}
which result from a coherent interaction of the resonance with
a nonresonant background.
In a previous paper \cite{roxana} an S-matrix analysis of these
line shapes was derived:
the resonance corresponds to a pole in the S-matrix 
in the 'unphysical' sheet
of the complex energy plane
\cite{bohm}. A systematic linearization of the slowly varying
parts of the S-matrix \cite{roxana} in the vicinity of this 
pole yields an expansion
for the relevant real energies close to the center $E_1$ of the resonance
as given by
\begin{equation}
S(E) = S_{res}  \frac{i{\Gamma \over 2}}  {E - E_1 + i \Gamma/2}
+ S_{bg} =  S_{bg} \frac{ \epsilon + q} {\epsilon + i}.
\label{fano}
\end{equation}
Here $S_{res} = S(E_1) - S_{bg}$ and $e = 2 (E-E_1) / \Gamma$,
where $\Gamma$ is the width of the resonance. Since the
pole strength $S_{res}$ and the background transmission
coefficient $S_{bg}$ are independent from each other
it is immediately clear that the asymmetry parameter
$q = i S(E_1)/S_{bg}$ is complex in the generic case.

Applying the Landauer-B\"uttiker formalism
one obtains from Eq.\ (\ref{fano})
Fano-type resonances in the stationary conductance \cite{roxana}. 
In real experiments one has to add to this
coherent conductance a noncoherent conductance background
resulting from independent transmission channels
(see, for example, 
Eq.\ (3) in Ref.\ \cite{goe2} where a real $q$ was assumed).
These extra channels
might either be coherent channels that do not couple to the
resonant channel or independent incoherent transport channels.
As a basic problem, it is known \cite{abstreiter,pan,roxana}, 
that such a noncoherent conductance background
has the same effect on the conductance
than a coherent background term caused by
$S_{bg}$.
Therefore, in a  fit of the stationary conductance
with a general Fano profile 
a coherent conductance background $S_{bg}$
cannot be resolved from a noncoherent conductance background.
In consequence it is not possible to extract the parameters
$q$ and $S_{bg}$ uniquely from a measurement of the
stationary conductance.

To  overcome the described problem of a stationary
conductance measurement we analyze the pulse transmission
properties in presence of a Fano resonance. 
Assuming a simple Gaussian incident
pulse and the validity of Eq.\ (\ref{fano}) we obtain
a first analytical description to cover
the pulse transmission in the entire Fano resonance regime.
Furthermore, in difference to previous calculations on resonant
pulse transmission
\cite{villa,konsek,pereyra1,calderon1,grossel,pereyra2,mizuta,harada,stovneng}
in our model the properties of the transmitted pulse can be
related directly to the critical parameters $q$ and $S_{bg}$.
We demonstrate that
in a pulse transmission experiment,
one can distinguish between a resonant and a
nonresonant conductance background
making it possible to determine the parameters $q$ and $S_{bg}$
uniquely.
However, 
at typical confinement lengths in the order of tens of nanometers
we find typical times to resolve the needed shape of
the transmitted pulses in the order 
of picoseconds.  Such a time regime is 
very hard to achieve experimentally and we are not aware of any
of such data. 

In a system without background  transmission
our analytical results  are in agreement with numerical calculations
by Konsek and Pearsall \cite{konsek}.
Neglecting the coherent background
term in Eq.\ (\ref{fano})
our model allows for a continuous transition between
the limits of a broad and a narrow resonance
on scale of the energetic width of the incident packet:
in the first case a weakly distorted Gaussian (WDG)
transmitted pulse results
with a small delay
with respect to the unscattered pulse \cite{levin}.
In the limit of a narrow resonance considered by
\cite{konsek} we find that the transmitted signal is dominated
by the decaying state (DS) associated with the pole in Eq.\ (\ref{fano}).
This  DS corresponds to a solution
of the stationary Schr\"odinger Equation with purely
outgoing boundary condition and with the complex eigenenergy
$E_1 - i \Gamma/2$ \cite{perelomov,stovneng,wagner}. It is
found that for narrow resonances
the dropping flank of the transmitted
pulse is not Gaussian any more but dominated
by a slow exponential time characteristic
$\propto \exp{(-\Gamma/2t)}$ associated with the complex
part of the eigenenergy.
In addition,
there appear characteristic oscillations in space and time
in the absolute
value of the wave function of the transmitted pulse.

\begin{figure}[b]
\begin{center}
\includegraphics*[width=3.in]{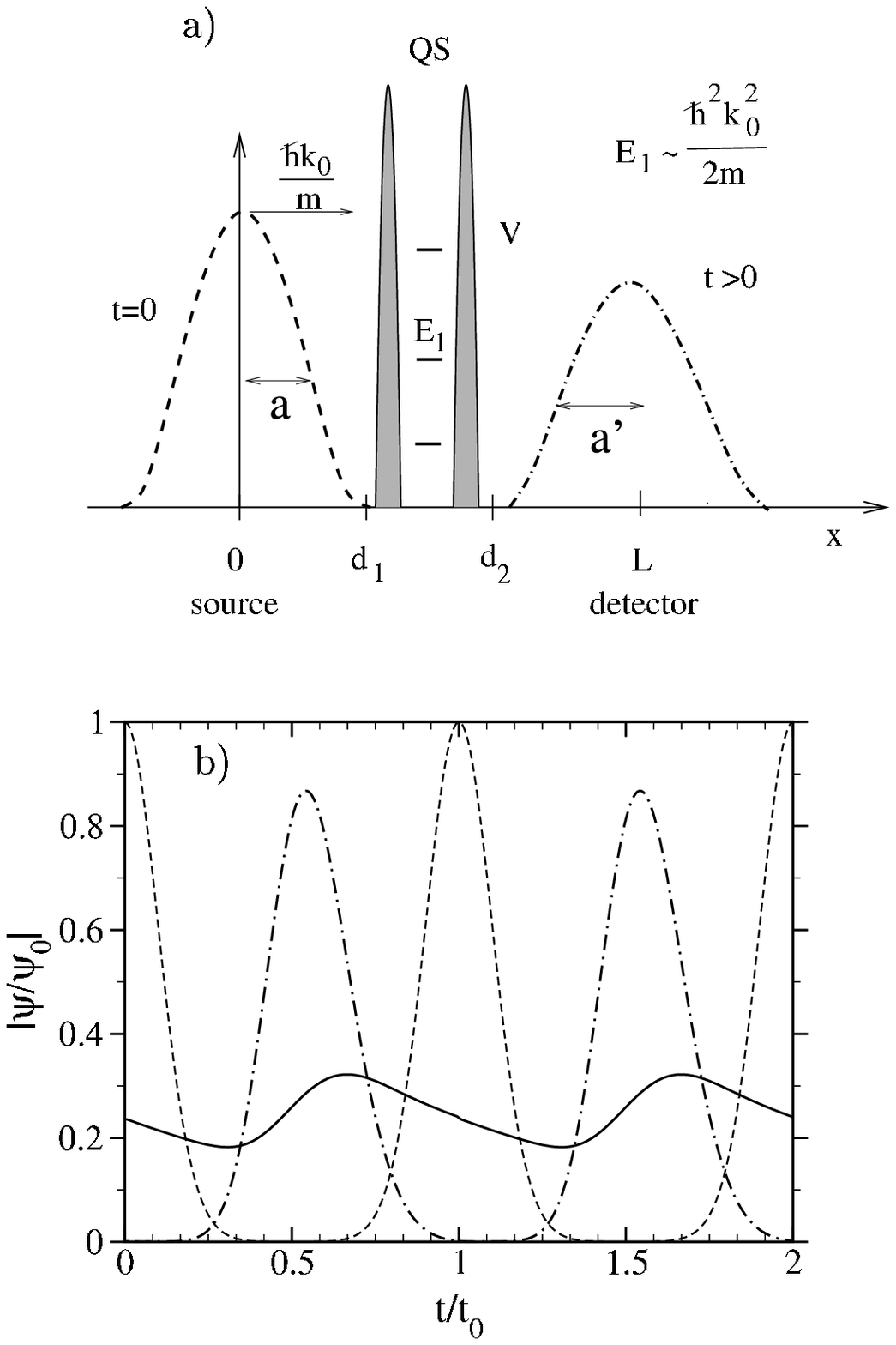}
\end{center}
\caption{ Part (a):
schematic representation of the incident Gaussian packet (dashed line)
approaching the quantum system.
The transmitted signal (dash-dotted line) is detected at 
$x=L$. \\
Part (b): absolute value of the wave function versus time. 
In dashed line the sequence of the incident 
pulses at $x = 0$. 
In dash-dotted line
the quasi-stationary limit for the transmitted pulses detected at $x=L$
for $\rho=2$ (broad resonance) and in solid  line for $\rho = 0.1$
(narrow resonance). 
The further parameters are $L = 5 a$, 
$k_0 a= k_1 a = 10$, and $|S_0|=1$.}
\label{Fig0}
\end{figure}

\section{model}
We analyze a set-up as depicted in Fig.\ \ref{Fig0}:
an  incident standard Gaussian wave packet 
\begin{equation}
\psi_0(x,t) =
{\psi_0 \over \sqrt{1 + i{\hbar \over m a^2 } t }}
\exp{ \left[ - {(x-v_0t)^2 \over  2 a^2 (1 + i {\hbar \over m a^2} t) } \right]}
\exp{\left[i k_0 x - i {\hbar k_0^2 \over 2m} t \right]} 
\label{psio}
\end{equation}
is approaching  a tunneling barrier with a group velocity
 $v_0 = \hbar k_0/m$. At $t=0$ the incident packet has a width of $a$
in real space.
The transmitted pulse is given by the expression
\begin{equation}
\psi(x > d_2  ,t) = \int_0^{\infty} {dk \over \sqrt{2\pi} }  \psi_0(k)
S(k) \exp{ \left[i kx -  i {\hbar \over 2m} k^2 t \right]},
\label{genera}
\end{equation}
where $\psi_0(k)$ is the Fourier transform of $\psi_0(x,t=0)$.
A short range scattering potential is assumed 
so that $V(x)=0$ outside the interval $d_1 \leq x \leq d_2$.
Furthermore it is assumed that there is
an isolated  resonance at $ E_1 \approx \hbar^2k_0^2 /(2m) $,
i. e. close to the mean kinetic energy of the incident pulse.
Then in the range of finite $|\psi_0 (k)|$ the transmission coefficient
can be approximated by an expression 
\begin{equation}
S(k)= S_0 { i \frac{\Gamma_k}{2}  \over k - k_1 + i \frac{\Gamma_k}{2}}
+S_{b},
\label{sk}
\end{equation}
with  $S_0 =   S(k_1) - S_{b}$. With the usual transformation
$E = (\hbar^2 /2m) k^2$ one
can derive Eq.\ (\ref{sk}) from Eq.\ (\ref{fano}).
Here one has to take into account that the 
representation of the S-matrix in the complex energy plane
in Eq.\ (\ref{fano}) only holds for the lower part
of the unphysical branch of the energy plane 
including the real axis, i. e. the
imaginary part of the energy must be smaller than or equal to 
zero. In the complex k-plane this means $Re(k) \geq 0$ and
$Im(k) \leq 0$ so that the entire integration range in 
Eq.\ (\ref{genera}) is covered.
Comparing the denominators in the first factor on the right hand side
of  (\ref{sk}) and (\ref{fano}) 
one obtains the relations
$k_1 = \sqrt{2mE_1}/\hbar$ and $\Gamma_k = (\hbar v_1) ^{-1}  \Gamma$, with
$v_1 = \hbar k_1/m$ for $\Gamma / E_1 << 1$.
Furthermore it results that
\begin{equation}
S_0 = {\Gamma \over \Gamma_k} {m \over \hbar^2}
{1 \over k_1 - i \Gamma_k/2} S_{res},
\end{equation}
and
\begin{equation}
S_{b} = S_{bg} - {S_0 \over 2} {i \Gamma_k/2  \over k_1 - i \Gamma_k/2}.
\end{equation}
We define an asymmetry parameter of the Fano distribution in k-space
as given by $q_k = i S(k_1) / S_b$.
Combining Eqs.\ (\ref{genera}) and (\ref{sk}) an integral is obtained that
can be solved analytically \cite{grad},
\begin{equation}
\psi(x > 0 ,t)  = S_{0} \psi_0(x,t)
 \left[ - {1 \over (1 + i q_k)} +
 \rho \sqrt{\pi \beta} {\cal F} (i z)  \right].
\label{centra}
\end{equation}
Sometimes the function ${\cal F}$ with
\begin{equation}
{\cal F} (i z)  =  
 \exp{ (z^2) } \mbox{erfc} (z) 
\label{calf}
\end{equation}
 is  called Fadeeva function
where 
$\mbox{erfc}$ is the complementary error function.
The argument $z$ is given by
\begin{equation}
z = \left( {q \over 2 \sqrt{\beta }}
-i \gamma \sqrt{\beta} \right),
\label{zvalue}
\end{equation}
with
$q = (x - v_0 t)/a$, $\beta = (1 + i \tau)/2$, $\tau = t/ t_0$,
$t_0 =  m a^2/\hbar$,
and  $\gamma = a[k_0 - k_1 + i\Gamma_k /2 ] \equiv
\Delta + i \rho $. The first factor in the square bracket
in Eq.\ (\ref{centra})
results from the transmission background $S_b$ in Eq.\ (\ref{sk}),
the second term results from the resonant term in  (\ref{sk}).

When the background term is neglected
it is possible to formally relate Eq.\ (\ref{centra}) 
to the basic problem of the propagation
of a step function modulated sine-signal in a dispersive medium
\cite{brillouin,buettiker,muga}. 
This model, initiated by Sommerfeld, is often invoked to discus general
features of signal transport.
As an example we consider Ref.\ \cite{buettiker}
where a dispersion of $\omega = 1 + k^2$ has been assumed.
Because of the structural equivalence
of Eq.\ (A1) of Ref.\ \cite{buettiker} 
and Eq.\ (\ref{genera})
the following mapping  between the basic quantities results: 
the variable
$a$ in  Eq.\ (A1) which is the time variable has to be identified
with $\beta$ in the present problem which is
essentially the complex time.
The variable $b$ in  Eq.\ (A1) which is the negative 
space coordinate has to be identified with $q$ which is a mixture
of space and time. Finally, the wave vector $k_0$ in Eq.\ (A1)
which is purely  imaginary for an evanescent medium 
(frequency in the gap) and
purely real in a propagating medium (frequency in the band)
has to be identified 
with $\gamma$. A systematic analysis of
the relation between the present model
and the Sommerfeld model using the saddle
point method is left to a subsequent paper.
Here we only emphasize that in the Sommerfeld model
the whole signal transfer process is regarded to take place in the
same medium which is represented by a complex refraction index.
In our case there is a clear distinction
between the sender domain, the transmission structure,
and the receiver domain. As illustrated in Fig.\ \ref{Fig0}
this distinction
is inspired by the possible use of a resonant
tunneling diode as a signal transmission structure, where
the sender domain is provided by the source contact
and the receiver domain by the drain contact.

\section{Resonance without background transmission}

We begin our analysis considering systems where the background
transmission can be neglected, $|S_b| \ll |S(k_1)|$ and therefore
$|q_k| \rightarrow \infty$. Then Eq.\ (\ref{centra}) becomes
\begin{equation}
\psi(x > 0 ,t)  = S_{0} \psi_0(x,t) \rho \sqrt{\pi \beta} {\cal F} (i z).
\label{centred}
\end{equation}
From (\ref{centred}) it 
follows that the wave function as a function of $q$ and $\tau$
depends only on the complex parameter $\gamma$
apart from the global transmission coefficient
$S_0$.  In this paper $\gamma = i\rho$
is chosen purely imaginary.
Depending on the parameter $\rho$ 
two different pulse propagation regimes 
result from Eq.\ (\ref{centred}).
They are illustrated in 
Fig. \ \ref{Fig0}(b).  Here
we plot the transmitted signal in the stationary state
resulting from  a sequence of 
well separated incident Gaussian pulses which are created at
integer $t/t_0=n$ with the center in real space at $x=0$.
For the broad resonance
($\rho >1$) the transmitted pulse at $x = L$ is a 
sequence of well separated WDG pulses.
These pulses arrive
with a small delay with respect to the times $t/t_0 = n + 1/2$ 
at which the
signal without scattering potential would arrive.
In contrast, if the  resonance is  narrow ($\rho \ll 1$)
 the
transmitted pulses are strongly weakened and  deformed so that they are
not separable any more. We will show below the reason
for this drastic degradation of the signal transfer: 
for a narrow resonance the signal is dominated by the slow
exponential time  characteristic of the  DS with a
life time that exceeds the time interval of two subsequent
pulses. 

The parameters used for the calculations in Fig.\ \ref{Fig0}(b) have been 
chosen to demonstrate that a nearly optimal signal transfer rate
can be achieved for a broad resonance. 
We call a signal transfer rate (number of detected pulses at $x = L$
per time) optimal under the following conditions:
first, the input pulses follow each other in a minimum
time interval to allow for a separation at $x=0$,
second, the  output pulses are still well separable
 at $x=L$, and, third,
the attenuation of the pulses is low
when passing from $x=0$ to  $x=L$.
To achieve this goal the following design rules have been applied:
i.) the detector at  $x=L$  is positioned in a minimum distance of
$L \approx 5 \div 10 a$ from the source at $x=0$. 
As illustrated in Fig.\ \ref{Fig0}(a) this choice
results from the requirements, first, that when the pulse is prepared
or detected it should be well separated from
the resonant tunneling structure and, second, that the spatial width of
the signal is assumed to be comparable  in size to the 
tunneling structure. ii.) the traveling time of the signal
which is the time the maximum of a given pulse needs to pass $x=L$
should not exceed the characteristic time $t_0 = ma^2 /\hbar $
of the quantum spread 
of the free wave packet in Eq.\ (\ref{psio}).
For an electron packet with a width of $a=10nm$ this time
is very short, $t_0 \approx 10^{-12}s$. 
We can therefore define a minimum 
speed $v_{min} = L / t_0 = \hbar  k_{0;min} /m$
which defines a minimum central position $k_{0;min}$ of the Gaussian
peak in $k$-space
(we neglect here the spread
induced by the Coulomb interaction). iii.)
for a maximum signal intensity
we require $k_0 = k_1$ 
($\Delta=0$) which is  optimal resonance.
iv.) the imaginary part of the resonance energy is fixed by the
condition $\rho > 1$ with the constraint that the resonance
should still be  isolated. 
v.) for an effectively symmetrical
double barrier structure the absolute value of $S_0$ is very close to
unity \cite{paul}. Apart from iv.) 
these rules are met
in the on-resonance calculations in Ref.\ \cite{konsek}.
Here we find $\rho \sim 0.1 <1$ as in the system
represented by solid lines
in  Fig.\ \ref{Fig0}(b).

\begin{figure}[h]
\begin{center}
\noindent \includegraphics*[width=3.7in]{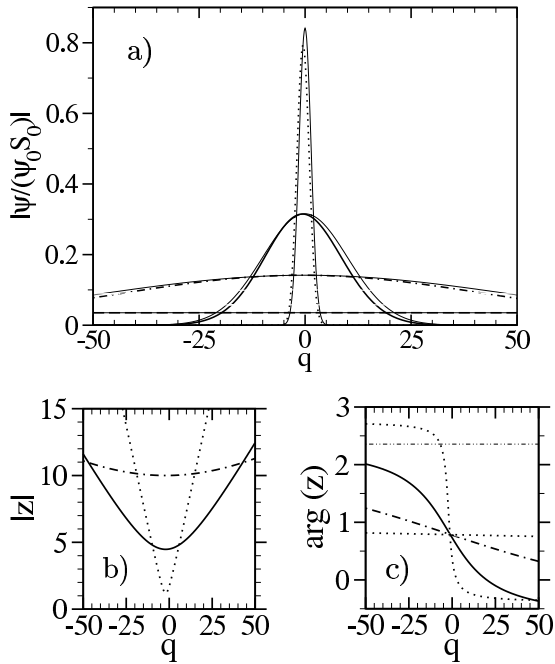}
\end{center}
\caption{Part a): Absolute value of the transmitted wave function  for
a broad resonance
 with $\rho =2.0$ at $\tau = 1.0$ (dotted line),
 $\tau = 10$ (solid line), $\tau =50$ (dash-dotted line),
and $\tau = 800$ (dotted line). Thick lines represent the exact result
in Eq.\ (\protect\ref{centra})
and, for comparison,
thin solid lines the result for $\psi = \psi_0(x,t)$
in Eq.\ (\protect\ref{psio}).
For $\tau = 800$ the corresponding
lines are not resolvable.
Part b) absolute value of $z$
(see Eq.\ (\protect \ref{zvalue})) and
part c) arg($z$) with the same line-coding as in part a).
The  thin dash-dot-dotted
line in $(c)$ represents $\mbox{arg}(z) = 3\pi/4$.}
\label{Fig2}
\end{figure}

\begin{figure}[b]
\begin{center}
\noindent \includegraphics*[width=3.5in]{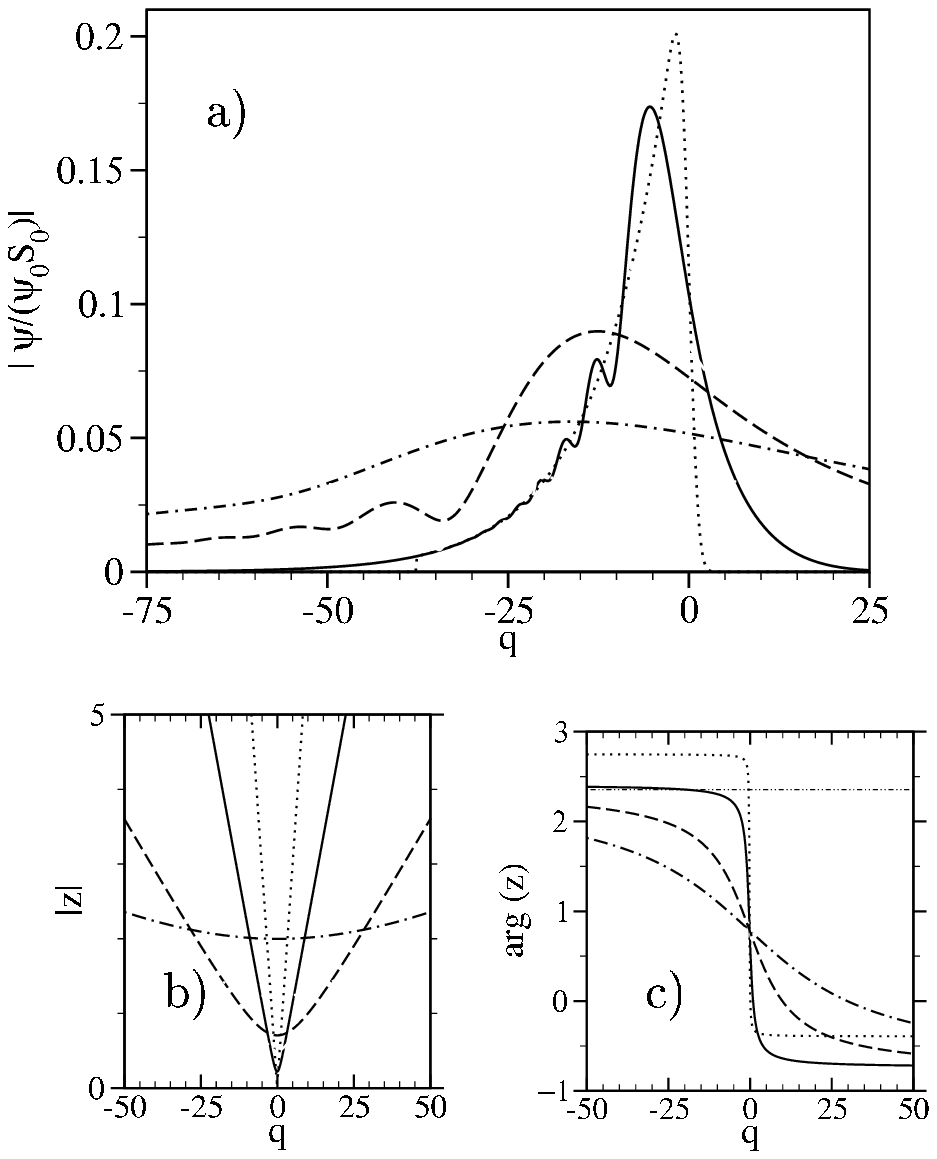}
\end{center}
\caption{(a) Absolute value of the transmitted wave function,
(b) absolute value of $z$,  and (c) the argument of $z$
at $\rho =.1$, $l_1 = l_0$ $\tau = 1.0$ (dotted line),
and
 $\tau = 10.$ (solid line), $\tau = 100$ (dashed line),
and $\tau = 300$ (dash-dotted line). The  thin dash-dot-dotted
line in $c$ represents $\mbox{arg}(z) = 3\pi/4$.}
\label{Fig3}
\end{figure}

A detailed comparison of the pulse propagation properties
in the regime of broad resonances and in the regime of narrow resonances
is contained in  Fig.\  \ref{Fig2} and Fig.\ \ref{Fig3}.
We will demonstrate that the basic differences can be 
understood in a proper expansion of the complementary 
error function in Eq.\ (\ref{centred}). As shown in Fig.\  \ref{Fig2}
for a broad resonance  Eq.\ (\ref{centred}) describes a 
WDG pulse. In this case we find
in the relevant q-range
(q values for which $|\psi|$ deviates significantly from zero)
that  the absolute value of $z$ is relatively
large and that the argument of $z$ is less than $3\pi/4$ (see
Figs.\ \ref{Fig2} (b) and (c)).
Under this condition
we can introduce an asymptotic $z$ expansion of the error function 
as given in \cite{abramowitz}  yielding
\begin{equation}
\psi(x,t) =
S_0 {\rho \sqrt{\beta} \over z}
\left[1 +\sum_{m=1} {a_m \over (2z^2)^m} \right]
\psi_0(x,t) \equiv \psi_{WDG}(x,t),
\label{appro1}
\end{equation}
with $a_m = (-1)^m [1 \times 3 \times 5  ...\times(2m-1)]$.
The expression in Eq.\ (\ref{appro1}) 
clearly reveals the WDG character of the transmitted pulse:
the weak distortion of the incident
Gaussian pulse $\psi_0(x,t)$ follows, first, from the fact that at
large $\rho$
the parameter $1/(2|z|^2)$ is small compared to unity. 
Therefore, the
second term in the square bracket in Eq.\ (\ref{appro1})
gives only a small correction.
In fact, only the first two terms in the sum over $m$ in  
Eq.\ (\ref{appro1}) are necessary
to approximate the analytical result from Eq.\ (\ref{centred})
within plot resolution.
Second, for large $\rho$ the argument $z$ depends only very
weakly on $q$. This  explains 
that the factor $1/z$ in front of the square
bracket in Eq.\ (\ref{appro1}) merely 
leads to  a minor deformation of the packet as well.
Note that in the illustrated example $\rho =2$ takes a moderate
value and that the approximation in Eq.\ (\ref{appro1}) 
works even better for larger $\rho$ for a small amount of terms
in the $m$ summation.

An inspection of Fig.\ \ref{Fig3} (a) shows immediately
that for a narrow resonance
the unperturbed incident Gaussian pulse is no good approximation for the
transmitted pulse. The difference is most pronounced for short
times. For example, at  $\tau =1$ a narrow Gaussian 
transmitted peak results in the case of a broad resonance
(Fig.\ \ref{Fig2} (a)) while for the narrow resonance
(Fig.\ \ref{Fig3} (a))  a  much weaker, broader, 
and strongly asymmetric resonance results which 
is nearly completely restricted to negative $q$.
To explain the difference we observe that 
for the broad resonance the
argument of $z$ is smaller than  $3\pi/4$
in the relevant q-range of the transmitted peak
($-5 \leq q \leq 5$ at $\tau =1$,
see Figs.\ \ref{Fig2} (b) and (c)). In contrast,
for the narrow resonance  
the argument  of $z$ is larger than $3\pi/4$ 
in the relevant q-range of the transmitted peak
($q \leq 0$ see  Figs.\ \ref{Fig3} (a) and (c)).
To obtain for small $\rho$ an approximate expression for
$q \leq 0$
one expands $\mbox{erfc} (-z)$ according to \cite{abramowitz}
instead of  $\mbox{erfc} (z)$ as has been
done for the broad resonance.
Then, a superposition of {\em two} factors
\begin{equation}
\psi(x,t) = \psi_{WDG} (x,t) + \psi_{DS}(x,t),
\label{rep}
\end{equation}
is obtained with
\begin{equation}
\psi_{DS}(x,t>0)  =  \psi_0 S_0  \rho
\sqrt{2\pi} \exp{ (\beta \rho^2)}
\exp{ \left[ {\Gamma_k \over 2} (x-v_0 t) \right] } 
\exp{\left(i k_0 x - i {\hbar k_0^2 \over 2m} t \right)}.
\label{decay}
\end{equation}
For the narrow resonance the factor $\psi_{DS}(x,t>0)$ 
is the dominant contribution to the signal and $\psi_{WDG} (x,t)$
is small.
The dominant factor $\psi_{DS}$ results
from a DS. This can be seen, first, from
the exponential decay $\propto \exp{ (-\Gamma t /2) }$ 
with $\Gamma = v_0 \Gamma_k$
at a fixed space coordinate $x$. 
Second,
at a fixed time the wave function grows $\propto \exp{ (\Gamma_k x/2) }$
which is typical for a DS \cite{perelomov}.
This growth continues up to $q \approx 0$ where the wave front
of the decaying state signal is located. As can be gathered from
Fig.\ \ref{Fig3} (b) this wave front 
can well be described by replacing in Eq.\ (\ref{centra})
the complementary error function by its small $z$-expansion
$1 - 2z/\sqrt{\pi}+...$\cite{abramowitz}.

\begin{figure}[t]
\begin{center}
\noindent \includegraphics*[width=3.0in]{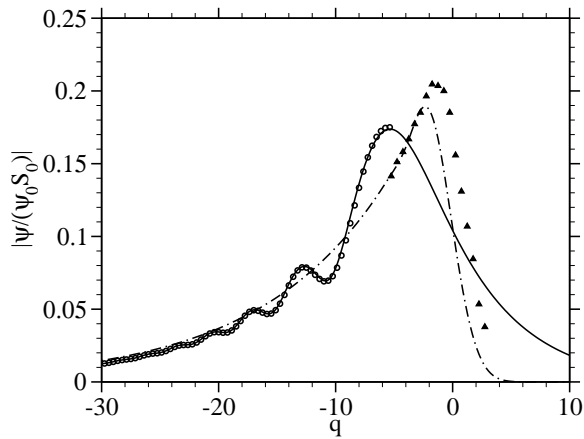}
\end{center}
\caption{
Solid line exact result for $\rho =0.1$ and $\tau =10$ (see Fig.\ (3))
and in
in open circles approximation $|\psi| = |\psi_{DS}| + \psi_{OSZ}$
with $ \psi_{OSZ}$ as given in Eq.\ (\ref{final}).
Dash-dotted line exact result for $\rho =0.092$ and $\tau =1.44$
and triangles numerical results for this case taken from  Ref.\ \cite{konsek}
(see text).
For the representation of the numerical results
we set $q = x - x_0 -v_0 t$, where
$x_0 = 50nm$ is the position of the center of the incident 
numerical packet at $t=0$.}
 \label{Fig4}
 \end{figure}

In Fig.\ \ref{Fig3} (a) the curve for  $\tau= 10$ 
shows characteristic
oscillations  superimposed to the
decaying state signal which  are presented in
more detail in Fig.\ \ref{Fig4}. 
For $q \leq -8$ the exact result
from Eq.\ (\ref{centred})
agrees within plot resolution
with the approximations in  Eqs.\ (\ref{appro1}) - (\ref{decay})
(the sum over $m$ in Eq.\ (\ref{appro1})
can be neglected).
The latter equations demonstrate the origin of the 
oscillations. They are caused
by the interference between the
dominant  component $\psi_{DS}$ 
and the small component $\psi_{WDG}$.
Here the oscillating
phase is in the Gaussian factor  $\exp{[ - q^2/(4  \beta ) ]}$
in $\psi_{WDG}$ at complex $\beta$. To show this
we introduce further approximations in Eq. \ (\ref{appro1}): 
first, for moderate times $\tau \sim 10$ one
may set $\beta \sim i \tau /2$
outside the
Gaussian factor  $\exp{[ - q^2/(4  \beta ) ]}$ and, second,
for small $\rho$ the contribution $\rho \sqrt{\beta}$ in $z$
can be omitted.  It results that
\begin{equation}
\psi_{WDG} = S_0 \psi_0 {\sqrt{\tau} \over q}  \rho
\exp{ \left(-{q^2 \over 4\beta} + i {\pi \over 4} \right)}
\exp{\left[i k_0 x - i {\hbar k_0^2 \over 2m} t \right]}.
\label{small}
\end{equation}
With this approximation for $\psi_{WDG}$ and with Eq.\ (\ref{decay})
a representation
$|\psi(x,t)| = |\psi_{DS}(x,t)| + \psi_{OSZ}(x,t)$ follows
where
\begin{equation}
\psi_{OSZ} = {\mbox{Re} (\psi_{DG} \psi_{DS}^*) \over | \psi_{DS}|} 
 =
 - |\psi_0 S_0| \rho { \sqrt{\tau} \over q} 
\exp{ \left( - {q^2 \over 2 (1 + \tau^2)} \right)}
\sin{ \left( {q^2 \over 2\tau} - {\tau \over 2 } \rho^2 - {\pi \over 4}
      \right)}.
\label{final}
\end{equation}
As shown in in Fig.\ \ref{Fig4} this expression
describes the characteristic oscillations very accurately.
Apart from the constant phase $-\pi /4$
the argument in the sine-factor in Eq.\ (\ref{final})
stems from $\exp{[ - q^2/(4  \beta ) ]}$ in Eq.\ (\ref{small}).
This argument determines solely
the  phase and the period of the characteristic oscillations.

In Fig.\ \ref{Fig4} we represent numerical data taken from Konsek
and Pearsall  for the 'on-resonance' wave packet  as given in Fig.\ (3)
of Ref.  \cite{konsek}.
To include these data in Fig.\ \ref{Fig4} 
we read off from Fig.\  2 (a) in Ref.  \cite{konsek}
the values $a=2nm$ and $|\psi_0| = \sqrt{28\times 10^{-3}} = 0.167$
for the incident Gaussian packet.
From Fig.\  2 (b) we estimate for the width 
of the resonance the value
$\Gamma = 25meV$.
For the 'on-resonance' wave packet ($\Delta = 0$)
the center of the resonance
is given by $E_0 = \hbar^2k_0^2/(2m) $, where $\hbar k_0 /m = v_0$
and $v_0 = 4.1 \times 10^5 ms^{-1}$ is the speed of the 
unscattered Gaussian package. We find 
$\rho = a \Gamma_k/2 = a \Gamma / (2 \hbar v_0) = 0.092$. After
having been prepared at $t=0$ the
shape of the wave packet is determined numerically
at a time $t = 5 \times 10^{-14}$
so that $\tau = 1.44$. The 
parameter pair $\rho =0.092$ and  $\tau = 1.44$
is actually very close to the parameter pair
$\rho =0.1$ and  $\tau = 1.0$ for which the exact result
is plotted in Fig.\ \ref{Fig3} (dotted line).  
It follows immediately that we can describe the
transmitted wave packet in Ref.\ \cite{konsek}
with Eq.\ (\ref{rep}). 
Because of the small time $\tau$
one has a dominant DS component (Eq.\ (\ref{decay}))
while the WDG
component in the transmitted signal is nearly negligible.
Then no characteristic oscillations as in  Eq.\ (\ref{final})
appear.  
The small differences between our analytical result and 
the numerical result by Konsek and Pearsall can be
attributed to imprecisions in our determination of $\Gamma$ and
to deviations of the numerical line shape from the Breit-Wigner profile.
One source for such a deviation
is the existence of a background transmission. 
The effect of a  background transmission will be discussed in the next section.

\section{Resonance with a background transmission }

\begin{figure}[bh]
\noindent\includegraphics*[width=3.in]{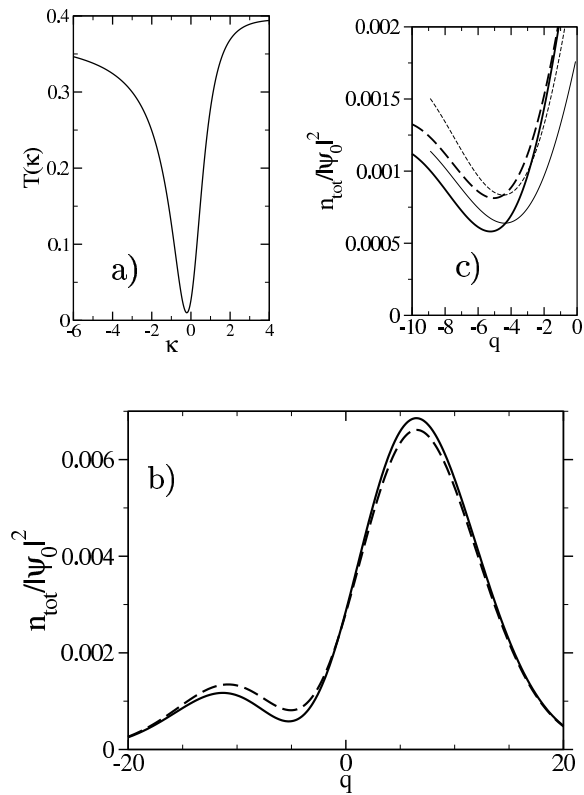}
\caption{a) Identical static transmission 
 for $T_b = 0.38$, $T_{of} =0$,
$Re(q_k)= 0.195$, $Im(q_k)= 0.165$ [$\alpha = 1$, Eq.\ (\ref{genfa})]
and for $T^F_b = 0.37$, $T^F_{of} = 0.01$, $q_F = 0.2$
[$\alpha =0$, Eq.\ (\ref{genrel})].
b) Total transmitted pulse with
the two resonances in a): 
$\alpha = 1$ (solid line, $|S_0|^2 = 0.28$)
and $\alpha = 0$ (dashed line, $|S_0|^2 = 0.385$) at $\tau =10$,
$\rho =2$, $\Delta =0$.
c) Detail of b). Thin lines
approximation in
Eq.\ (\ref{broadback}).}
\label{stami}
\end{figure}

In this section we want to demonstrate that in the presence
of a  background transmission a pulse transmission experiment can
yield valuable information about the scattering potential
beyond the measurement of the stationary conductance.
To be specific  we focus on the case of an anti-resonance with
$|q_k|\ll 1$.
Furthermore, the anti-resonance
is assumed to be broad compared to the incident Gaussian
pulse, $\rho \geq 1$. 
We begin our analysis
considering the static transmission calculated from Eq.\ (\ref{sk}) 
which is is given by
\begin{equation}
T(k) = |S(k)|^2 = T_b {[\kappa+Re(q_k)]^2 + Im(q_k)^2 \over \kappa^2 +1}
+ T_{of},
\label{genfa}
\end{equation} 
with $\kappa = (k - k_1)/(\Gamma_k/2)$ and $T_b = |S_b|^2$.
The static transmission of a typical anti-resonance
resulting from Eq.\ (\ref{genfa}) for $T_{of} =0$
is  illustrated in Fig.\ \ref{stami} (a). There is
a pronounced dip  for small
$|\kappa|$
which produces a transmission 
zero at $\kappa = -q_k$ for real $q_k$.
If a small
imaginary part is introduced in $q_k$
the transmission shows only a minimum, i. e.
there is a residual transmission close to the position
of the  transmission zero at real $q_k$.
Like in  Ref. \cite{goe2} (see Eq.\ (3) therein)
we added in Eq.\ (\ref{genfa}) a constant background term $T_{of}$ 
to take into account the contribution of nonresonant
independent transmission channels. These channels
might either be coherent channels that do not couple to the 
resonant channel or independent incoherent transport channels.
We therefore call $T_{of}$ a noncoherent conductance background
(noncoherent with respect to the resonant channel).
However, in difference to Ref. \cite{goe2} the 
coherent contribution in  Eq.\ (\ref{genfa})
which causes the resonance is represented with a complex asymmetry 
parameter $q_k$. In Ref. \cite{roxana} we have shown
that such a complex asymmetry parameter $q_k$ is the generic
result for a general system. 
Now, if one applies Eq.\ (\ref{genfa})  to the interpretation 
of experimental conductance peaks there arises a central problem:
as illustrated
in Fig.\ \ref{stami}  there is a one-dimensional manifold
of choices for the four parameters
$T_b,  T_{of}, Re(q_k), Im(q_k)$ leading to the same
function $T(k)$ in Eq.\ (\ref{genfa}). 
Therefore, if only $T(k)$ is determined
by measuring the static conductance only three independent
parameters can be determined in a fit procedure.
We choose these parameters as
$T_b^F$, $T_{of}^F$, and $q_F$ with real $q_F$ 
which result from a fit of the expression
\begin{equation}
T(k) = T^F_b {(\kappa+ q_F)^2 \over \kappa^2 +1} + T^F_{of}.
\label{genrel}
\end{equation}
Such an expression was used in Ref. \cite{goe2} to
fit Fano-type conductance resonances in quantum dots
in the strong coupling regime. However, 
at complex  $q_k$ the factor $T^F_{of}$
cannot be taken as the contribution of the noncoherent
channels. This is demonstrated
if  one inserts in Eq.\ (\ref{genfa}) the transformation
\begin{equation}
T_b = T_b^F + \alpha T^F_{of},
\label{tra1}
\end{equation}
\begin{equation}
Re(q_k) = q_F {T_b^F \over T_b^F + \alpha T^F_{of}},
\label{tra2}
\end{equation}
\begin{equation}
Im(q_k) = {T_{of}^F \over T_b^F + \alpha T^F_{of}}
\sqrt{\alpha^2 + (1+q_F^2) \alpha {T^F_b \over T^F_{of}}},
\label{tra3}
\end{equation}
and
\begin{equation}
T_{of} = (1- \alpha) T_{of}^F,
\label{tra4}
\end{equation}
with the real parameter $\alpha$ chosen freely within the interval
$0\leq \alpha \leq 1$. This transformation
leaves the function $T(k)$ invariant. For $\alpha = 1$
one recovers Eq.\ (\ref{genfa}) and for $\alpha = 0$
one obtains Eq.\ (\ref{genrel}).
From Eq.\ (\ref{tra4}) it follows that 
any measured fit parameter $T_{of}^F$ is compatible
with zero contribution of noncoherent channels,
$T_{of} =0$, if one takes $\alpha = 1$. Then the measured
minimum conductance originates completely from the
complex part of the asymmetry parameter.
On the other hand a measured fit parameter $T_{of}^F$ is
maximally compatible with a  contribution of noncoherent channels
as given by $T_{of} =T_{of}^F$. In this special case
the asymmetry parameter is real and $\alpha = 1$.

In the following we want to demonstrate that in principle
it is possible to determine $\alpha$ 
with a combination of experiments on 
pulse transmission and  on static transport.
Here a measurement of $\alpha \neq 1$ would provide
evidence for the correctness of Eq.\ (\ref{genfa}).
In Fig.\ \ref{stami} (b) results for the density $n_{tot}$
of the total transmitted pulse 
are plotted with
\begin{equation}
|n_{tot}(x,t)|^2 =  T_{of}   |\psi_0(x,t)|^2 + |\psi(x,t)|^2.
\label{compo}
\end{equation}
Here the first factor 
arises from the constant noncoherent 
background transmission of the incident pulse $\psi_0(x,t)$.
The second factor is given by the independent coherent transmission
of the pulse via the resonant channel with a static 
transmission coefficient as given by
Eq.\ (\ref{sk}). The wave function intensity coming 
from the coherent transmission component
is the absolute square of Eq.\ (\ref{centra}).
The parameters $|S_0|^2, |\psi_0|^2, q_k, \rho$, and $\tau$
which enter in the absolute square
of Eq.\ (\ref{centra}) as a function of $q$ we obtain
in the following way:
first, we assume that from a fit of the experimental static conductance
according to Eq.\ (\ref{genrel}) the parameters $k_1$ and $\Gamma_k$
are known so that the incident packet can be 
prepared, for example,
at $\Delta =0$ and $\rho =2 >1$ (broad resonance, see Fig.\ \ref{stami}). 
Second,  from the fit of the static conductance the parameters
$T_b^F$, $T_{of}^F$ and $q_F$ can be found. These three parameters
fix according to Eqs.\ (\ref{tra2}) and  (\ref{tra3})
a range of possible $T_b$, $T_{of}$, and complex $q_k$ which is parametrized by $\alpha$.
Then, it follows easily that
\begin{equation}
|S_0|^2 = T_b \left\{ [1 - Im(q_k)]^2 + Re(q_k)^2] \right\}.
\label{s02}
\end{equation}
Third, to keep the analysis simple we consider in our theory
the transmitted packet at a constant time
which we choose to be relatively short,
$\tau =10$, so that particular features are not washed out
in the time evolution.
Fourth, we observe
that both factors on the right hand side of Eq.\ (\ref{compo})
are proportional to the square to the parameter
$|\psi_0|$ so it can be accounted for as a normalization parameter.
In Fig.\ \ref{stami}(b) we compare the transmitted pulse 
calculated from Eq.\ (\ref{compo}) for $\alpha =1$ and for $\alpha =0$.
As a qualitatively new feature 
we find that in both cases there appears
a pronounced dip
in $n_{tot}$ close to the center the pulse.
As demonstrated in Fig.\ \ref{stami}(c) 
this dip  can be explained inserting in Eq.\ (\ref{compo})
\begin{equation}
\psi(x,t) = S_0 \psi_0(x,t)  i \left[ q_k + {q \over  \rho \tau} \right],
\label{broadback}
\end{equation}
valid for $|q/(\rho \beta)|\ll 1$. 
The expression in Eq.\ (\ref{broadback}) can be derived
from Eq.\ (\ref{centra}) approximating in the first factor in the
square bracket $1/(iq_k +1) \approx 1 - iq_k$ and introducing
in the second factor Eq.\ (\ref{appro1})
where we have dropped the qualitatively irrelevant summation over $m$.
Finally like for Eq.\ (\ref{final}) we have assumed moderate times
so that $\beta \approx i \tau /2$. From Eq.\ (\ref{broadback})
a minimum in $n_{tot}$ 
follows at the position $q = -  \rho \tau Re(q_k)$. 
As can be seen from Figs.\ \ref{stami}(b) and (c) there are
differences in the transmitted pulse at $\alpha =1$
and $\alpha =0$ making a determination of $\alpha$
possible. These differences are most pronounced in the region of
the dip: the minimum of the absolute square of the 
wave function associated with the dip is about 25\%
smaller for $\alpha =1$ than for $\alpha=0$. Because 
of the spatial propagation of the transmitted pulses
this effect also exists when the pulse is measured
as a function of time in an appropriate position in space.
Therefore, in principle,
with a  measurement of the minimum of the transmitted pulse
associated with the dip it is possible to determine $\alpha$.
A pulse measurement 
can therefore resolve uncertainties in $\alpha$
resulting in a sole measurement of the static conductance.

\section{Conclusions}

We have analyzed an analytical model for the resonant transmission
of a Gaussian wave packet. 
At zero background transmission and in the limit of a broad
resonance a weakly distorted transmitted pulse results.
In the opposite limit of a narrow resonance the transmitted wave
is dominated by
a decaying state and characteristic oscillations result.
Our model also describes pulse transmission in presence of a
coherent and a noncoherent transmission background. 
In this case the transmission is described by a general Fano profile.
We focus
on the case of an anti-resonance. We demonstrate
that the characteristic effect of the transmission background
is the formation of dips in the profile of the transmitted
pulse.
It is shown that the measurement of the pulse transmission 
in addition to static transport opens the
possibility to distinguish between a coherent and a noncoherent
conductance background. This distinction is clearest when
the dip in the pulse profile is considered.

We acknowledge helpful discussions with D. Robaschik and P. Racec.

\end{document}